\documentclass[final,5p,times,twocolumn]{elsarticle} 
\usepackage{amssymb,amsmath}
\usepackage{tabularx}
\usepackage[table,xcdraw]{xcolor}
\usepackage{natbib}
\usepackage[greek,english]{babel}
\usepackage{subcaption}
\usepackage{float}
\usepackage[colorlinks=true,linkcolor=blue,urlcolor=blue]{hyperref}
\usepackage{ulem}

\makeatletter
\def\ps@pprintTitle{%
 \let\@oddhead\@empty
 \let\@evenhead\@empty
 \let\@oddfoot\@empty
 \let\@evenfoot\@oddfoot
}
\makeatother

\begin{document}

\begin{frontmatter} 
\title{Limiting risk to reduce inequality: insights from the Yard-Sale model}
\author[lauti]{Lautaro Giordano\corref{cor1}}
\ead{lautaro.giordano@ib.edu.ar}
\cortext[cor1]{Corresponding author}

\author[nacho]{Ignacio Cortés}
\ead{ignaciocortes@gmail.com}

\author[seba]{Sebastian Gonçalves}
\ead{sgonc@if.ufrgs.br}

\author[fabi]{María Fabiana Laguna}
\ead{lagunaf@cab.cnea.gov.ar}

\address[lauti]{Statistical and Interdisciplinary Physics Group, Centro Atómico Bariloche (CNEA) and CONICET. Instituto Balseiro, Universidad Nacional de Cuyo. R8402AGP Bariloche, Argentina.}

\address[nacho]{Statistical and Interdisciplinary Physics Group, Centro Atómico Bariloche (CNEA). R8402AGP Bariloche, Argentina.}

\address[seba]{Instituto de Física, Universidade Federal do Rio Grande do Sul, 91501-970 Porto Alegre RS, Brazil.}

\address[fabi]{Statistical and Interdisciplinary Physics Group, Centro Atómico Bariloche (CNEA) and CONICET. R8402AGP Bariloche, Argentina.}

\begin{abstract} 
Wealth inequality remains a critical socioeconomic challenge, driven by systemic dynamics and self-reinforcing mechanisms that amplify the economic imbalances. Simplified models from statistical physics provide valuable insights into the fundamental mechanisms governing wealth distribution. In this study,  we extend the Yard-Sale model ---a minimal kinetic exchange framework--- to investigate how limiting risk in economic transactions affects inequality. While previous research demonstrates that such models naturally lead to wealth concentration, we introduce a mechanism that restricts the maximum risk agents can assume during exchanges. Numerical simulations reveal that this modification fosters more equitable wealth distributions and significantly reduces extreme disparities. These findings highlight the importance of individual-level constraints in shaping systemic outcomes, offering new perspectives on promoting economic balance.
\end{abstract}

\begin{keyword}
Econophysics, Wealth Distribution, Inequality, Risk
\end{keyword}

\date{\today}
\end{frontmatter}

\section{Introduction} \label{sec:intro}
Economic inequality and wealth distribution are central topics in socioeconomic research, reflecting both fundamental human capital factors, such as education and health, and broader systemic dynamics. While the mechanisms behind wealth inequality are complex, they often create feedback loops that reinforce disparities, making it essential to unravel their causes and effects. 
In particular, Piketty extensively analyzed the concentration of wealth among elites \cite{PIKETTY2014}, highlighting the role of inherited wealth and unequal returns as key factors driving inequality in modern economies.
However, inequality can spontaneously emerge even in the absence of inheritance or mechanisms favoring the wealthy.

Studying the distribution of wealth in a society requires taking into account a myriad of factors. Economies encompass diverse productive activities, markets, and a wide range of heterogeneous agents. In addition, the sources of income that drive inequality are varied and multifaceted.
Although capturing the full complexity of these processes is beyond the scope of simplified models, theoretical frameworks from statistical physics provide valuable tools for investigating key aspects of wealth exchange~\cite{PATRIARCA2010}. 
In particular, many models focus on random pairwise exchanges of wealth, where wealth is usually conserved. It is also common to assume that agents save a fraction of their resources.~\cite{CHAKRABORTI2000}. 
A prominent example is the Yard-Sale model, which, despite its simplicity, provides insights into the dynamics of wealth exchange and serves as the basis for this work~\cite{HAYES2002,JULI1}.

Previous studies, both numerical~\cite{CAON2006} and analytical~\cite{MOUKARZEL2007, BOGHOSIAN2015, CARDOSO2021}, have shown that kinetic exchange models, including more general pairwise interaction models~\cite{CARDOSO2023}, consistently converge towards a state of wealth concentration.
From the perspective of microscopic models in Econophysics, this result is particularly noteworthy as the dynamics of unbiased wealth exchange naturally leads to concentration, even in the absence of pre-existing disparities in assets or productive capacities. To mitigate this undesirable outcome, various modifications to the basic models have been proposed. These include mechanisms that favor less wealthy agents during transactions~\cite{IGLESIAS2004, SCAFETTA2002}, the introduction of taxation and redistribution schemes~\cite{LI2019, BUSTOS2016, LIMA2020, IGLESIAS2020}, or the incorporation of a finite fraction of random exchanges into the otherwise strict Yard-Sale model~\cite{GHOSH2023}. In this work, we adopt the first approach, focusing on strategies to limit wealth concentration.
Specifically, we investigate how constraining risk during economic exchanges within the Yard-Sale model impacts wealth distribution. By limiting the maximum risk agents can assume, we analyze the emergence of more equitable wealth distributions and the reduction of extreme disparities. This approach provides a fresh perspective on the dynamics of wealth inequality by emphasizing the interplay between individual-level interactions and systemic constraints, offering insights into the mechanisms that foster greater economic balance.

In the next section (Sec.~\ref{sec:model}), we present the model along with the specific control parameters, the measures, and the technical details of the numerical simulations. This is followed by the presentation of the results (Sec.~\ref{sec:results}) and, finally, the conclusions (Sec.~\ref{sec:conclusions}).

\section{Model} \label{sec:model}

The Yard-Sale model considers a system of $N$ interacting agents, each characterized by its wealth $w$ and a risk factor $r$. The risk factor defines the fraction of wealth an agent is willing to risk in each transaction and remains constant throughout the simulation. This parameter is directly related to the marginal propensity to save $\beta$, commonly used in kinetic exchange models, through the relation $r = 1 - \beta$~\cite{CHAKRABORTI2000}. In fact, as discussed in the \textit{Conclusions} section, this quantity is closely related to Keynes’ marginal propensity to consume, which provides additional economic motivation for its use in the model. At each Monte Carlo step (MCS), all agents are iterated and for each agent $i$, a random partner $j$ is selected, ensuring that every agent interacts at least once. 
According to the Yard-Sale interaction rule, the wealth transferred from the losing agent to the winning agent is
\begin{equation}
    \Delta w_{ij} = \min(r_iw_i, r_jw_j).
    \label{eq:deltawij}
\end{equation}
In this model, both agents are always able to fully complete the exchange without the need to acquire debt ($w<0$). In addition, the total wealth of the system is conserved at each interaction. Agent $i$ is chosen as the winner with probability
\begin{equation}
    p_{ij} = \textrm{Prob}{(\textrm{$i$ wins})} = \frac{1}{2} + f \, \frac{w_j - w_i}{w_i + w_j},
    \label{eq:p_iwins}
\end{equation}
while $p_{ji} = 1 - p_{ij}$. The parameter $f \in [0, 0.5]$, referred to as the social protection factor~\cite{SCAFETTA2002, SCAFETTA2004}, biases the winning probability in favor of the poorer agent. This serves as a straightforward method to model state intervention aimed at mitigating wealth concentration without the need for explicit redistribution mechanisms. At $f = 0$, both agents have the same probability of winning the exchange, and we recover the original Yard-Sale model. In this limit, the model reaches a state of maximum inequality, with all wealth concentrated in a single agent~\cite{CAON2006, MOUKARZEL2007, BOGHOSIAN2015, CARDOSO2021}.

The initial wealth of the agents is chosen randomly from a normalized uniform distribution so that $W = \sum_{i} w_i(0) = 1$. This normalization has the advantage that each agent's wealth also represents its fraction of the total wealth. 
For computational purposes, the minimum wealth required for an agent to participate in a trade is set as $w_{min} \approx 10^{-17}$. This value is several orders of magnitude smaller than the mean wealth per agent and can be effectively considered to be virtually zero. 
Furthermore, the results presented in the next section do not show significant changes for small deviations from the chosen $w_{min}$ value.

Agents with wealth above $w_{min}$ are called \textit{active} agents. There is no redistribution mechanism, so once an agent becomes inactive, it cannot return to the active state. The fraction of active agents at a given time provides insight into the state of the system, as it measures the effective size of the group of agents exchanging wealth at that moment.

In previous works, the risks are all equal to a fixed value~\cite{IGLESIAS2004, CHAKRABORTI2002, VAZQUEZMONTEJO} or uniformly distributed in the range $[0,1]$~\cite{JULI1, PATRIARCA2013, BHFC2020}. In this work, we introduce a new parameter $r_{max}$ that limits the maximum possible risk, uniformly distributing the risks in the range $[0, r_{max}]$.

A widely used measure of inequality is the Gini index, defined as the relative mean absolute wealth difference between agents:
\begin{equation}
    G(t) = \frac{1}{2NW} \sum_{i,j} |w_i(t) - w_j(t)|.
    \label{eq:gini}
\end{equation}
The Gini index equals $0$ when all agents hold the same amount of wealth. At the other extreme, when the index reaches $1$, all the wealth in the system is in the hands of a single agent.

Unless otherwise indicated, the results presented in the next section correspond to an ensemble of $1000$ independent simulations with $N = 1000$ agents. There is one ensemble for each $r_{max}$. 
Simulations were run for $5\times 10^5\, \textrm{MCS}$, during which the Gini index stabilizes and changes by less than 0.5\% even if the simulation runtime is doubled, thereby defining the stationary state.

\section{Results} \label{sec:results}

\subsection*{Wealth distributions}

\begin{figure}[h]
\centering
\includegraphics[width=\columnwidth]{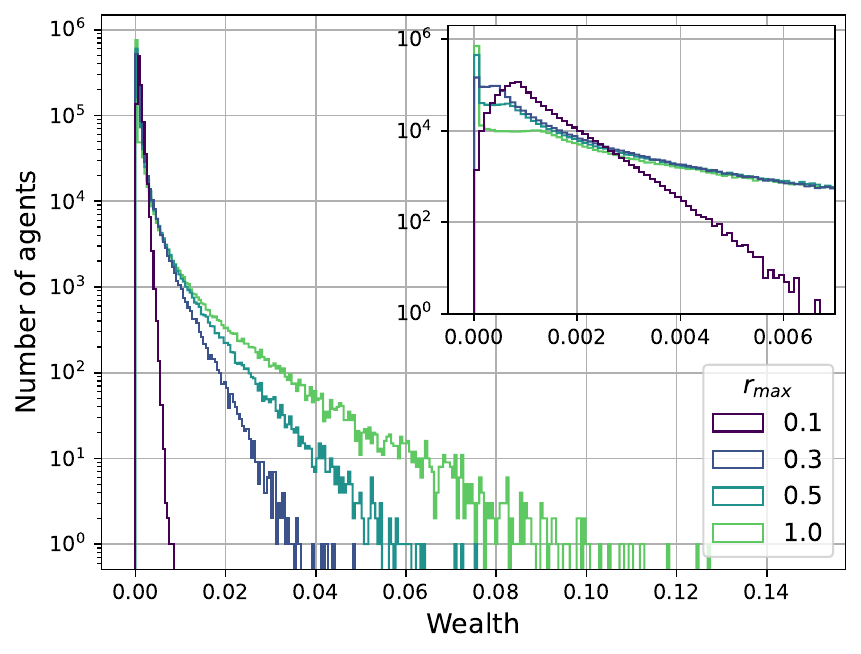}
\caption{Stationary wealth distributions for $f=0.1$ and four selected values of $r_{max}$ in semi-log scale, based on 1000 independent runs. The inset highlights variations in the distribution shapes within the lower wealth region.}
\label{fig:distr_riqueza_inset}
\end{figure}

In this section, we present the key results derived from the previously described model. Figure~\ref{fig:distr_riqueza_inset} illustrates the stationary wealth distributions for a low protection factor ($f=0.1$) and selected values of the maximum risk, including $r_{max}=1$, which corresponds to the Yard-Sale model studied in prior work~\cite{JULI1}.
In the high $r_{max}$ regime, the presence of wealthy agents comes at the expense of a significantly large number of agents holding near-zero wealth. As $r_{max}$ decreases, the distribution becomes tighter, with fewer agents concentrating large amounts of wealth. 
This is more clearly illustrated in the inset, which provides a close-up of the low-wealth region and shows that the frequency in the first bin increases as $r_{max}$ increases.

\begin{figure}[h]
\centering
\includegraphics[width=\columnwidth]{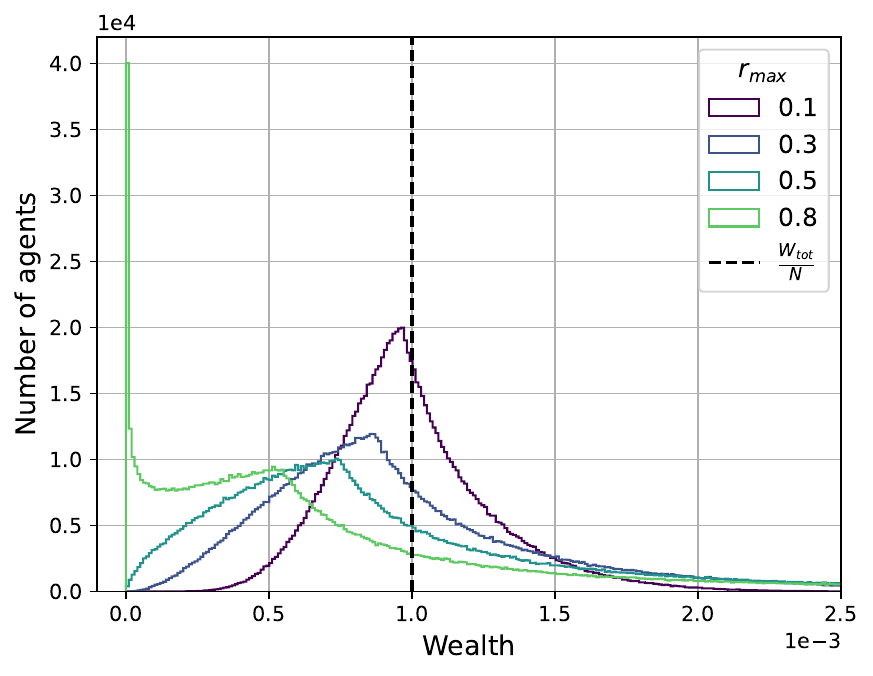}
\caption{Stationary wealth distributions for $f=0.3$ and four selected values of $r_{max}$ in linear scale, based on 1000 independent runs. The dashed line represents the mean wealth per agent in the system, which is $10^{-3}$ in this case.}
\label{fig:distr_riqueza_lineal}
\end{figure}

A similar trend is observed in the wealth distributions for all values $f > 0$. Fig~\ref{fig:distr_riqueza_lineal} illustrates the case for $f = 0.3$. As $r_{max}$ decreases, the distribution becomes increasingly concentrated around the mean wealth per agent (represented by a vertical dashed line), indicating that most agents possess a similar share of the total wealth. Notably, for $r_{max} \leq 0.5$, the number of very poor agents is drastically reduced.

\subsection*{Gini index}

\begin{figure}[h]
\centering
\includegraphics[width=\columnwidth]{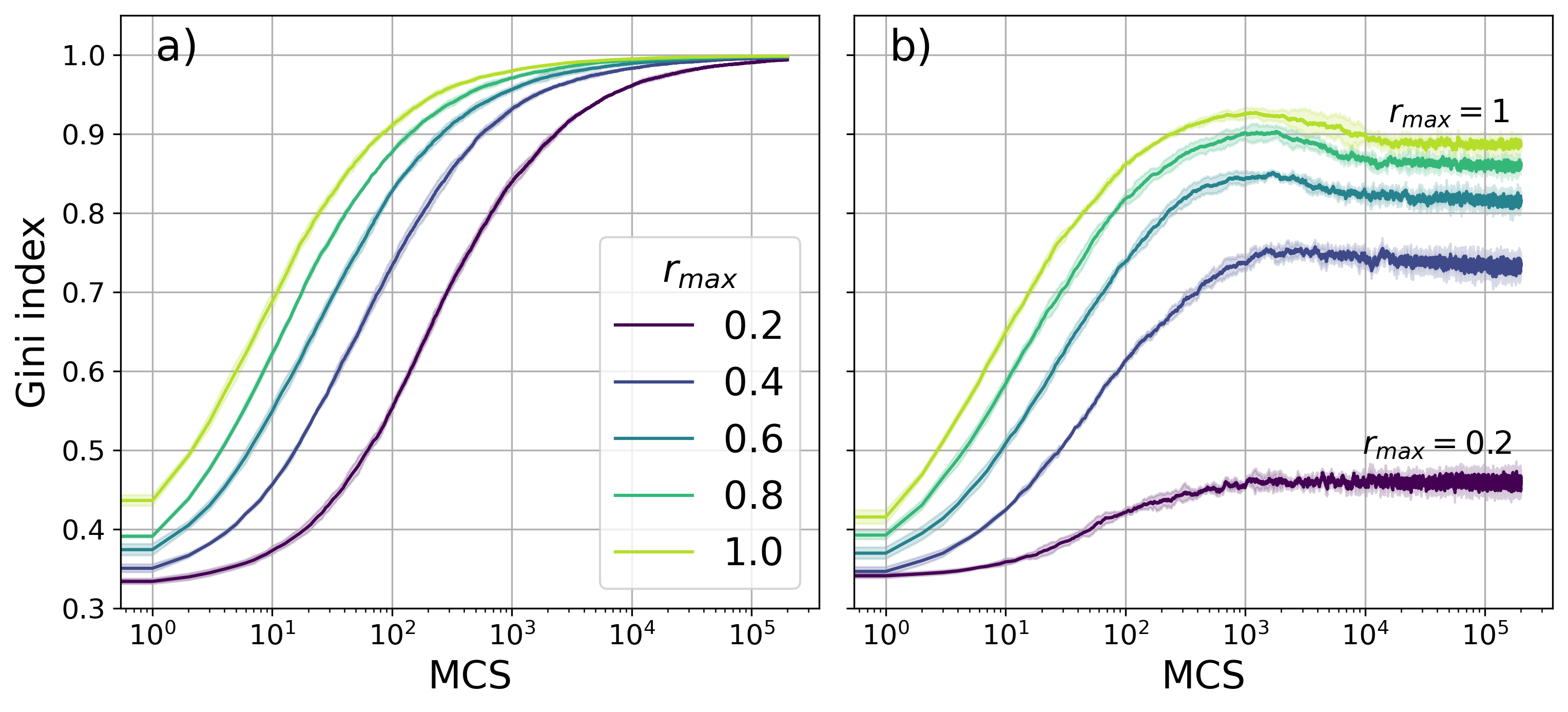}
\caption{Time evolution of the Gini index for different values of $r_{max}$ and two values of the protection factor: (a) $f=0$ and (b) $f=0.1$. Results are averages over 10 systems and the shaded region represents the standard deviation of the curves.}
\label{fig:dos_ginis}
\end{figure}

To better understand how the distributions are formed, we examine the temporal evolution of the Gini index, shown in Fig.~\ref{fig:dos_ginis} for $f = 0$ and $f = 0.1$, at several $r_{max}$ values, averaged over 10 independent systems per curve. For $f=0$, the system asymptotically approaches a Gini of $1$, indicating a state where a single agent accumulates all the wealth. This is the unavoidable limit of the Yard-Sale model for $f=0$, which has been extensively studied before~\cite{CAON2006, MOUKARZEL2007, BOGHOSIAN2015, CARDOSO2021}. The value of $r_{max}$ affects the time scale of wealth concentration, which is expected since decreasing the value of $r_{max}$ reduces the average wealth exchanged per transaction, requiring more interactions for the system to reach the concentrated state. In contrast, the case for $f = 0.1$ differs significantly, as the stationary value for the Gini index now depends on $r_{max}$. As $r_{max}$ decreases, the inequality in the system is reduced. Another notable difference between both cases is the amplitude of the fluctuations. For $f = 0$, a large number of agents are expelled in just a few steps, resulting in a smaller system of active agents with low-amplitude fluctuations. In contrast, higher values of $f$ allow more agents to remain in the system, leading to a noisier dynamics with larger amplitude fluctuations.

\begin{figure}[h]
\centering
\includegraphics[width=\columnwidth]{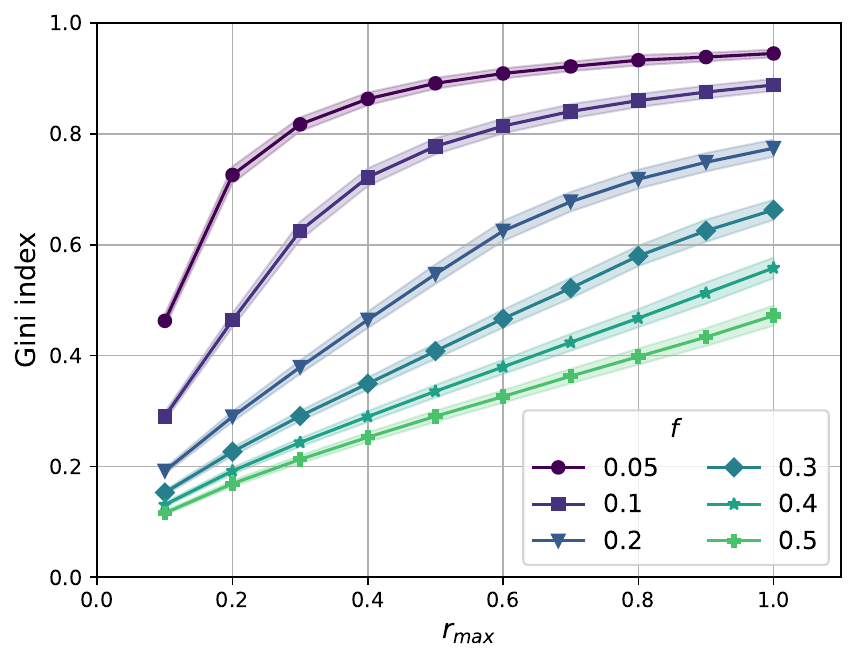}
\caption{Stationary Gini index as a function of $r_{max}$ for different values of $f$. The lines connecting the points serve as a guide for the eye, and the standard deviation of each point is represented by the shaded area between points.}
\label{fig:gini_final_muchos_f}
\end{figure}

The phenomenon of inequality reduction is observed for all values of $f > 0$, as shown in Fig.~\ref{fig:gini_final_muchos_f}, where the stationary Gini index values are plotted as functions of $r_{max}$ for various values of $f$. Across all curves, the Gini index increases with $r_{max}$, displaying a linear trend for high values of $f$ and low values of $r_{max}$. In addition, all curves exhibit a deviation from the linear behavior around a particular value, $r_{crit}$, which depends on $f$ (see next section). This deviation becomes more pronounced for higher values of  $f$, as $r_{crit}$ decreases with $f$, as will be shown shortly. A detailed analysis of this intriguing result is yet to be performed.

\subsection*{Relation between risk and wealth}

\begin{figure}[h]
\centering
\includegraphics[width=\columnwidth]{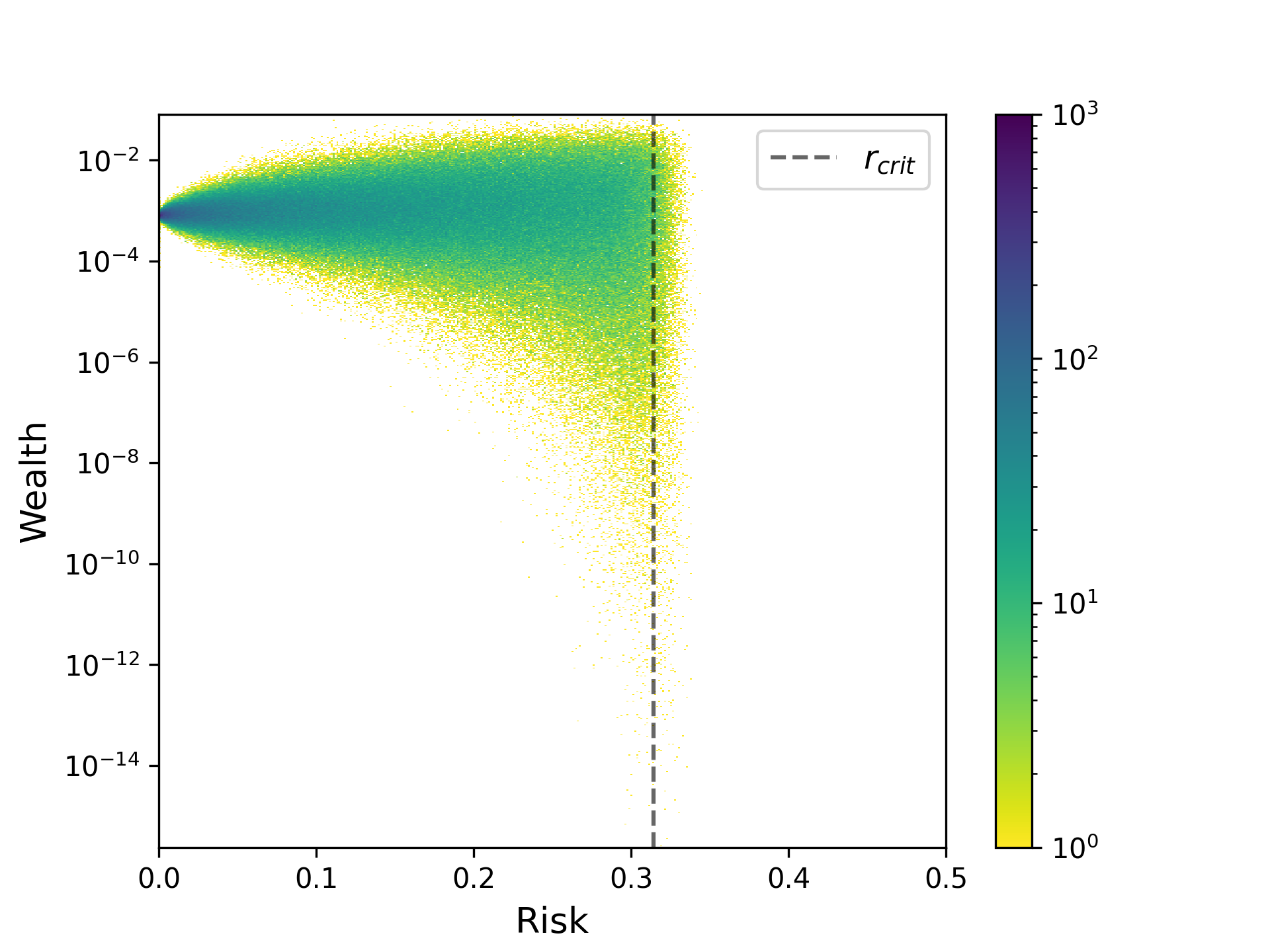}
   \caption{Density plot of agents wealth and risk for $1000$ independent systems with $f=0.1$ and $r_{max} = 0.6$ at $t = 5\times 10^5\, \textrm{MCS}$. Only active agents are shown for clarity. There is a limiting value $r_{crit} < r_{max}$, beyond which agents with higher risk lose all their wealth.}
    \label{fig:densityplot}
\end{figure}

To gain a deeper understanding of the effect of limiting risk, it is important to identify the characteristics of agents that accumulate significant wealth, as well as those of agents that tend to lose all their wealth, and how these results change with $f$. To do this, we analyze the relation between risk and wealth at the microscopic level. In Fig.~\ref{fig:densityplot}, we present a 2D plot showing the active agents density as a function of risk and wealth in the stationary state, for the system with $f=0.1$ and $r_{max}=0.6$. We observe that the spread on the final wealth of the agents increases with their risk $r$.
In addition, there is a risk value, $r_{crit}$, beyond which the number of active agents abruptly decreases to zero. This behavior was previously observed in~\cite{JULI1} for the original system with $r_{max}=1$. In the following section, we will provide a method to estimate the value of $r_{crit}$.

\begin{figure}[h]
\centering
\includegraphics[width=\columnwidth]{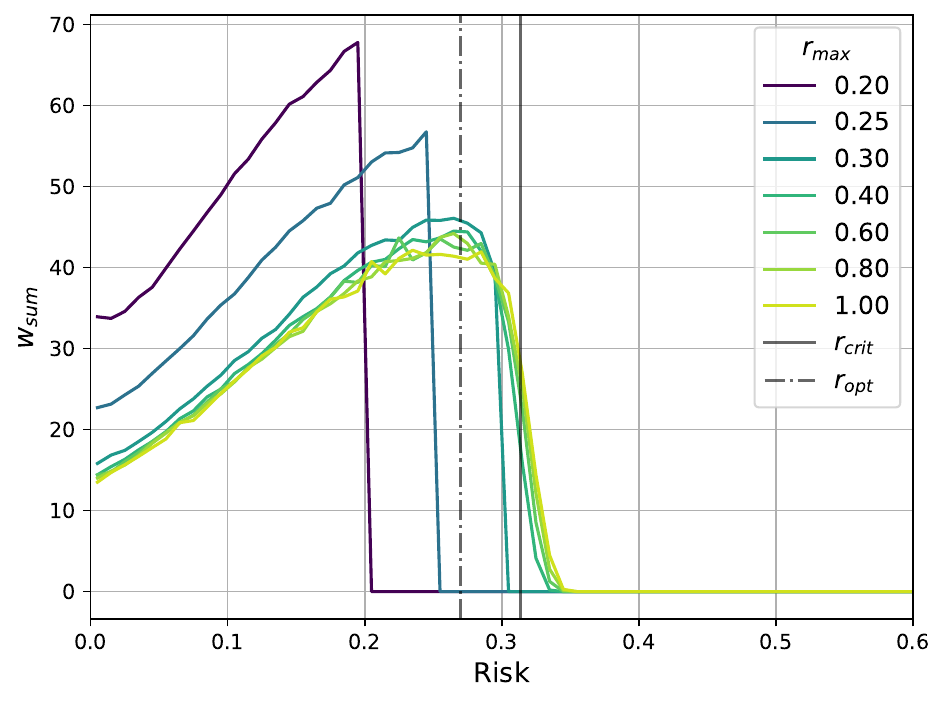}
   \caption{Wealth by bin of risk for $f = 0.1$. The total wealth of agents within each risk bin is plotted as a function of the maximum risk. Notice that for $r > r_{crit}$, all curves collapse onto a universal curve that shares the same optimal risk value, $r_{opt}$.}
    \label{fig:risk_bin_wsum}
\end{figure}

We extend our analysis to investigate how $r_{crit}$ changes over systems with different values of $r_{max}$. To achieve this, agents are grouped into risk bins, and the sum of their wealth is plotted as a function of that risk, as shown in Fig.~\ref{fig:risk_bin_wsum}. For low values of $r_{max}$, the curves increase monotonically, indicating that riskier agents concentrate more wealth than conservative ones. However, when $r_{max} > 0.3$, the curves converge to a universal shape, sharing the same limiting value, $r_{crit}$, and the same maximum wealth value, referred to as the optimal risk, $r_{opt} \approx 0.27$. The optimal risk, otherwise known as the $\textit{Kelly}$ risk~\cite{KELLY1956}, defines the optimal strategy an agent can assume in order to maximize it's earnings. Some work has been done in this direction~\cite{MOUKARZEL2011}, but a full analytical study of this problem is still pending, which may shed some light in understanding the inner mechanisms that govern these modified Yard-Sale models. 

The emergence of a universal curve is intriguing, because it occurs even though the number of agents per bin increases as $r_{max}$ decreases. Yet, the total wealth within each bin remains unaffected by this increase in agent numbers. This behavior can be understood by interpreting Fig.~\ref{fig:risk_bin_wsum} as a measure of the relative performance of agents with different risk levels. In the stationary state, any active agent $i$ with risk $r_i < r_{crit}$ interacts with equal probability with other active agents $j$ within the same risk range $(r_j < r_{crit})$. As a result, the total wealth in a bin reflects the relative performance of its agents compared to others. This performance measure is independent of $r_{max}$ because active agents remain uniformly distributed in the interval $[0, r_{crit}]$, even if the effective size of the systems varies with $r_{max}$. Moreover, since the total wealth is the same for all systems, the area under each curve must be identical regardless of $r_{max}$. Consequently, the curves collapse into the same shape, signifying that the distribution of wealth among risk levels remains unchanged for $r_{max} > r_{crit}$.

\subsection*{Estimation of $r_{crit}$}

The value of $r_{crit}$ for each $f$ can be estimated using different approaches. In this work, we demonstrate a new method to determine it by analyzing the fraction of active agents in the stationary state as a function of $r_{max}$. As previously noted, active agents are defined as those with wealth $w \geq w_{min}$. In this model, agents cannot regain active status once their wealth falls below this threshold in a transaction. 

For this calculation we will assume that $r_{crit}$ serves as a boundary that separates active agents from inactive ones. This assumption may be somewhat strong, as there are instances where inactive agents have risk values below $r_{crit}$, and active agents have risk values above it. A more accurate interpretation of $r_{crit}$ is as a threshold beyond which the probability of finding an active agent decreases significantly. 

Under the previous assumption, for a system with a uniform risk distribution between $0$ and $r_{max}$, if $r_{max} > r_{crit}$, the number of active agents will be proportional to $r_{crit}$. Conversely, if $r_{max} \leq r_{crit}$, all agents will be active.
Consequently, the fraction of active agents is described by

\begin{equation}
    n_{active}(r_{max}) = \min \left(1,\, \frac{r_{crit}}{r_{max}}\right).
\end{equation}

\begin{figure}[h]
\centering
\includegraphics[width=\columnwidth]{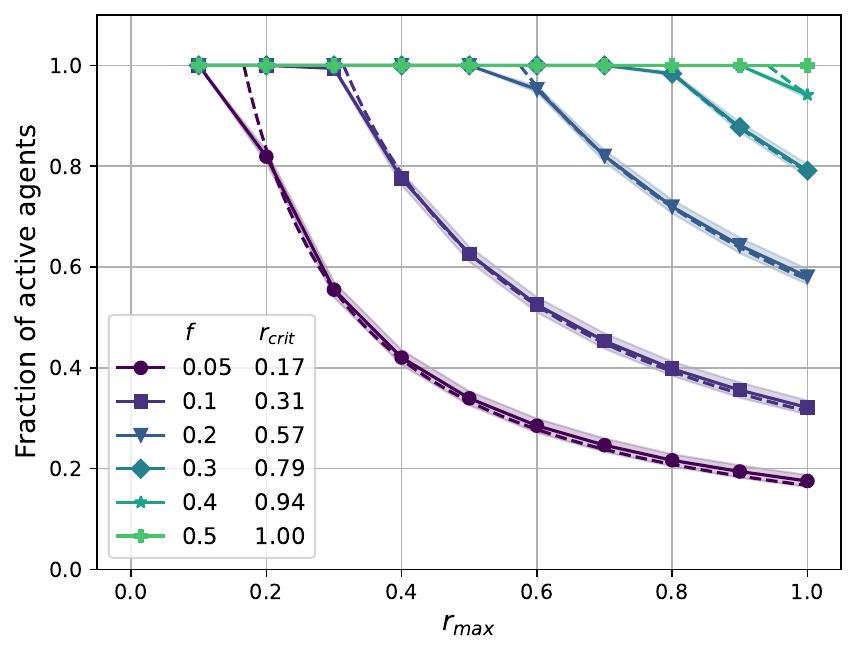}
\caption{Fraction of active agents at $t = 5\times 10^5\, \textrm{MCS}$ as a function of $r_{max}$ for several values of $f$.}
\label{fig:actives_final_muchos_f_fiteada}
\end{figure}

In Fig.~\ref{fig:actives_final_muchos_f_fiteada}, we show the fraction of active agents as a function of $r_{max}$ for several values of $f$, along with the fitted function $n_{active}(r_{max})$, from which the value of $r_{crit}$ can be estimated for each $f$.
We observe that the fraction of active agents decreases with $r_{max}$ for all values of $f$. This finding aligns with the analysis of the previous figures, as reducing $r_{max}$ increases the number of individuals in the system capable of participating in transactions, thus reducing inequality. We also observe that $r_{crit}$ increases with $f$, which means that a higher value of $f$ allows more risky agents to exist within the system.

\section{Conclusions} \label{sec:conclusions}
The analysis of wealth concentration has been a subject of interest in economic sciences since their inception. Aristotle already identified the concentration of wealth arising solely from transactional mechanisms as an activity ``against nature'' \cite{ARISTOTLE}.

The Yard-Sale model is one of the most widely used approaches to describe wealth distribution in societies due to its simplicity and effectiveness. The model relies on only two parameters: $f$, a general protection factor, and $r_i$, an individual parameter that limits the wealth exchanged in a single transaction. It assumes that all economic exchanges are represented as pairwise interactions within a mean-field framework, making it both versatile and, in some limited cases, analytically tractable.

In the present contribution, we use this model to investigate the impact of limiting the risk agents can assume during these pairwise exchanges. The rationale behind this proposal is that limiting the maximum risk in an exchange better aligns with the reality of transactions in society: individuals and companies often have fixed capital that they do not continuously wager in every economic transaction. Consequently, the risk taken by these agents should, in practice, be much lower than $1$. This problem has been widely studied by modern economists, including Keynes with his Marginal Propensity to Consume (MPC)~\cite{KEYNES}, which posits that as individuals' income increases, a smaller proportion of their total income is allocated to consumption, which can be directly related to the risk in our model.

Specifically, we examine how constraining the maximum risk and therefore reducing the average wealth exchanged per transaction affects the overall wealth distribution. The results, illustrated in Figs.~\ref{fig:distr_riqueza_inset} and~\ref{fig:distr_riqueza_lineal}, reveal that reducing the maximum risk leads to wealth distributions that are more concentrated around the system's average wealth. This shift reduces the number of agents with extremely high or very low wealth, promoting a more equitable distribution.

This behavior is consistent with the decrease in the Gini coefficient shown in Figs.~\ref{fig:dos_ginis} and ~\ref{fig:gini_final_muchos_f} as $r_{max}$ decreases across all systems with $f > 0$. The results indicate that for a fixed value of $f$, reducing $r_{max}$ leads to a significant reduction in inequality, which brings it closer to the levels observed in many countries around the world.  Although the Yard-Sale model specifically addresses wealth exchange, Gini index data are more commonly available for income distribution, reflecting its broader applicability in socioeconomic studies.

A comprehensive characterization of the model presented involved not only the study of macroscopic inequality indicators but also an analysis of the system's microscopic behavior as a function of $r_{max}$. 
We found that, as in the previously studied case of $r_{max} = 1$~\cite{JULI1}, a critical risk (defined as a risk value above which agents end up with no wealth) and an optimal risk (which corresponds to the risk value that maximizes the agent's wealth) exist, both dependent on $f$ but independent of $r_{max}$. 
This is clearly illustrated in Figs.~\ref{fig:densityplot} and~\ref{fig:risk_bin_wsum}.
In particular, Fig.~\ref{fig:risk_bin_wsum} shows that the curve of total wealth grouped by risk remains invariant with respect to $r_{max}$. This outcome is reasonable, as all systems with $r_{max} > r_{crit}$ share the same effective risk distribution after excluding inactive agents, even though the effective size of the system may differ. High-risk agents ($r_i > r_{crit}$) are doomed to lose their wealth as a consequence of the multiplicative dynamics of the model, which in turn increases the wealth held by low-risk agents, especially those near the optimal risk value. This phenomenon highlights the systemic consequences of allowing agents to take excessive risks. Their unavoidable losses tend to amplify disparities in a society and make the economy more unstable. Therefore, the existence of agents with extreme risk preferences can be viewed as a mechanism that fosters wealth concentration. 

Finally, we proposed a method to estimate $r_{crit}$ based on the fraction of active agents as a function of $r_{max}$, which can be observed in Fig.~\ref{fig:actives_final_muchos_f_fiteada}. Although this estimation depends on the definition of active agents (i.e., the minimum wealth threshold in the system), it is consistent with other measurements made in previous works using different methodologies \cite{JULI1, MOUKARZEL2007}.

Despite the simplicity of the modification we proposed for the Yard-Sale model, there are still open questions. One of them is understanding the behavior of the Gini index as a function of $r_{max}$ for different values of $f$ (see figure~\ref{fig:gini_final_muchos_f}). Although we found that the change in behavior in the curves occurs around the value of $r_{crit}$ corresponding to each $f$, a deeper understanding of the causes of this result is still needed. Another subject worth exploring is the stationary value of the total wealth transacted per Monte Carlo step as a function of $f$ and $r_{max}$, which may serve as a measure of the economy's efficiency. Preliminary results (not included here) suggest that this value does not exhibit monotonic behavior but rather reaches a maximum when $r_{max} \approx r_{crit}$.

Our findings demonstrate that incorporating individual-level constraints, such as limiting risk in economic transactions, can help reduce wealth concentration and promote greater economic balance, offering valuable insights for future research aimed at promoting a more equitable economic system.

\section*{Acknowledgements} \label{sec:acknowledgements}
S.G. acknowledges partial support from Conselho Nacional de Desenvolvimento Científico e Tecnológico (CNPq) of Brazil under grant \# 314738/2021–5.
The authors thank Thiago Dias, from the Universidade Técnica Federal do Paraná, Brazil, for fruitful discussions.


\end{document}